\documentclass[twocolumn,showpacs,preprintnumbers,pra,groupedaddress,amsmath,amssymb]{revtex4}

\usepackage{graphicx}% Include figure files
\usepackage[latin1]{inputenc}
\usepackage[usenames]{color}
\begin{document}

\title{Collective excitation of a Bose-Einstein condensate\\
    by modulation of the atomic scattering length}

\author{S. E. Pollack}
\author{D. Dries}
\author{R. G. Hulet}
\affiliation{Department of Physics and Astronomy and Rice Quantum
Institute,\\ Rice University, Houston, Texas 77005, USA}
\author{K. M. F. Magalh\~{a}es} 
\author{E. A. L. Henn}
\author{E. R. F. Ramos}
\author{M. A. Caracanhas}
\author{V. S. Bagnato}
\affiliation{Instituto de F\'{i}sica de S\~{a}o Carlos,
Universidade de S\~{a}o Paulo, Caixa Postal 369, 13560-970,
S\~{a}o Carlos-SP Brazil}

\date{\today}
%\date{January 12, 2010}

\begin{abstract}
We excite the lowest-lying quadrupole mode of a Bose-Einstein condensate
by modulating the atomic scattering length via a Feshbach resonance.
Excitation occurs at various modulation frequencies, 
and resonances located at the natural
quadrupole frequency of the condensate and at the first harmonic are observed.
We also investigate the amplitude of the excited mode as
a function of modulation depth.
Numerical simulations based on a variational calculation
agree with our experimental results and provide insight into
the observed behavior.
\end{abstract}


\maketitle

%%%%%%%%%%%%%%%%%%%%%%%%%%%%%%%%%%%%%%%%%%%%%%%%%%%%%%%%%%%%%%%%%%%%%
%\section{Introduction}
Collective excitation of a Bose-Einstein condensate (BEC)
is an essential diagnostic tool for investigating properties
of the ultracold quantum state.
Fundamental information about condensate dynamics
can be determined from observations of collective modes
\cite{jin96,mewes96,fort_euro}, 
including the effects of temperature \cite{jin97,stamper-kurn98,PhysRevLett.88.250402},
and the dimensionality of the system \cite{PhysRevA.72.053631}.
In addition, the interplay between these modes and
external agents, such as 
%permeable barriers \cite{engels:160405},
random potentials \cite{PhysRevLett.95.070401,disorder}, 
%\cite{falco:013624,PhysRevLett.95.070401,modugno:013606,nattermann:060402,disorder}, 
lattices \cite{PhysRevLett.90.140405,PhysRevLett.91.250402},
as well as other atoms 
\cite{PhysRevLett.81.1539, PhysRevLett.89.053202, PhysRevLett.89.190404, Ferlaino:joptB, mertes:190402}
%\cite{PhysRevLett.81.1539,PhysRevLett.87.080403,PhysRevLett.88.160401,PhysRevLett.89.190404}
can be investigated.
Examining the excitation spectrum of the BEC
allows for a detailed comparison 
with theoretical models \cite{PhysRevLett.77.1671,stringari96,jackson02,PhysRevLett.91.250403},
and related quantum systems such as 
superfluid helium \cite{PhysRev.79.309,adamenko:279} 
and superconductors \cite{PhysRev.115.795,PhysRevB.51.1147}.

Generically, a collective excitation is generated by the modification 
of the trapping potential of the condensate~\cite{RevModPhys.71.463}.
One convenient method is to 
apply a sudden magnetic field gradient, 
thereby shifting the center of the trap and exciting a 
dipole oscillation about the trap center.
One may also suddenly change 
the curvature of the trap to excite quadrupole modes.
The lowest-lying $m = 0$ quadrupole mode is characterized by 
out-of-phase axial and radial oscillations.

%%%%%%%%%%%%%%%%%%%%%%%%%%%%%%%%%%%%%%%%%%%%%%%%%%%%%%%%%%%%%%%%%%%%%%%
\begin{figure}[!b]
\includegraphics[width=1\columnwidth]{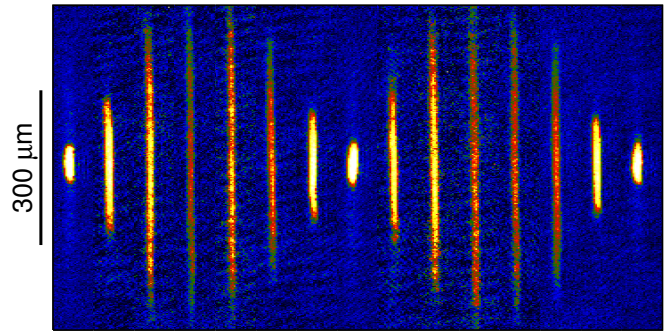}
\caption{ (color online)
    Quadrupole oscillation excited by the modulation of $a$.
    A series of \emph{in situ} polarization phase-contrast images of condensates 
    taken during excitation at $\Omega=(2\pi)\,10\,$Hz,
    separated in time by 15\,ms.
    The change in peak density is nearly an order of magnitude
    from the most compressed to the most extended condensate. 
    The horizontal scale of these images has been stretched by a factor of two for clarity.
    There is negligible excitation of the dipole mode.
\label{fig:image} }
\end{figure}
%%%%%%%%%%%%%%%%%%%%%%%%%%%%%%%%%%%%%%%%%%%%%%%%%%%%%%%%%%%%%%%%%%%%%%%%

If the condensate is not the only occupant of the trap, 
i.e., there exists a thermal component or another species of atoms,
then the other atoms may also be excited through these processes.
The evolution of a collective excitation can
therefore be complicated because the multiple components 
may affect damping or induce frequency shifts of the oscillation
\cite{PhysRevLett.81.1539, PhysRevLett.89.053202, PhysRevLett.89.190404, Ferlaino:joptB, mertes:190402}.
Therefore modulating the trap, although an
extremely useful tool for an isolated condensate,
can be cumbersome when the system to be studied is multi-specied.
An alternative approach is to excite the condensate alone, 
leaving the other occupants of the trap untouched.  

In this work, we demonstrate the excitation of the
lowest-lying quadrupole mode in a BEC of $^7$Li by modulating
the atomic scattering length via a magnetic Feshbach resonance.  
In contrast to abruptly changing the scattering length \cite{PhysRevLett.81.243},
sinusoidal modulation enables the controlled excitation of a single mode at
a specific frequency.
In addition, by using this method a coexisting thermal component
will be minimally excited by the mean-field coupling
to the normal gas \cite{PhysRevA.58.2423,meppelink-2009}.  
In the case of a multi-species experiment, 
resonant modulation of the scattering length
of one species will not necessarily excite
the others, depending on the details of other
Feshbach resonances present in the system \cite{frucg}.
Therefore, this technique may be useful for investigating
non-zero temperature effects, and as a powerful diagnostic tool for 
multi-species ultracold atomic experiments.
%\cite{ PhysRevLett.78.586, truscott, PhysRevLett.87.080403, PhysRevLett.88.160401, PhysRevLett.89.150403, PhysRevA.71.061401, PhysRevLett.95.170408, ospelkaus:020401, mcnamara:080404, taglieber:011402, pilch:042718, catani:011603, PhysRevLett.93.183201}.
%\cite{PhysRevLett.87.080403, PhysRevLett.88.160401,multiSpecied}.

%Consult: PRA 74:063601, Nature Phys 5:339 (2009), PRL 92:220403 (2004), PRL 86:2196 (2001), Rev Mod Phys 71:463 (1999).
%Temperature: PRA 71:043613 (2005), PRA 67:053607 (2003), PRL 85:4844 (2000), PRA 79:023602 (2009), PRL 88:180402 (2002), PRL 86:3938 (2001), PRA 62:053602 (2000), PRL 88:250402 (2002), and many others...

%%%%%%%%%%%%%%%%%%%%%%%%%%%%%%%%%%%%%%%%%%%%%%%%%%%%%%%%%%%%%%%%%%%%%%%%%%%%%
%%%%%%%%%%%%%%%%%%%%%%%%%%%%%%%%%%%%%%%%%%%%%%%%%%%%%%%%%%%%%%%%%%%%%%%%%%%%%
%\section{Collective Oscillations}
%%%%%%%%%%%%%%%%%%%%%%%%%%%%%%%%%%%%%%%%%%%%%%%%%%%%%%%%%%%%%%%%%%%%%%%%%%%%%

A trapped BEC at zero temperature may be described by the
three-dimensional cylindrically symmetric Gross-Pitaevskii equation \cite{pitaevskii}
\begin{equation}
i \hbar \frac{\partial}{\partial t} \psi = 
    -\frac{\hbar^2}{2m}\nabla^2 \psi
    + V \psi
    +\frac{4\pi \hbar^2 a}{m}\left|\psi\right|^2 \psi,
 \label{eq:gpe}
\end{equation}
where $m$ is the atomic mass,
the trapping potential is $V = (1/2) m (\omega_r^2 r^2 + \omega_z^2 z^2)$
with $\omega_z$ ($\omega_r$) the axial (radial) trapping frequency,
$a$ is the $s$-wave scattering length,
and the density is given by $n = |\psi|^2$.
It is convenient to introduce the anisotropy parameter
$\lambda = \omega_z / \omega_r$.
We use a variational approach to solve this equation and determine
the frequencies of the lowest-lying modes.
Using a Gaussian ansatz and minimizing the corresponding energy functional,
we derive the following coupled differential equations for the dimensionless 
axial and radial half-widths $u_z$ and $u_r$ of the condensate \cite{perez-garcia96}
\begin{equation}\label{eq:var}
\begin{split}
\ddot{u}_{r} + u_{r} &= \frac{1}{u_{r}^{3}} + \frac{P}{u_{r}^{3} u_{z}} \\
\ddot{u}_{z} + \lambda^{2} u_{z} &= \frac{1}{u_{z}^{3}} + \frac{P}{u_{r}^{2} u_{z}^{2}},
\end{split}
\end{equation}
where the interaction parameter $P = \sqrt{2/\pi} \left( N a/l_r\right)$,
with $l_r = \sqrt{\hbar / m \omega_r}$ the radial harmonic oscillator size,
and $N$ is the number of condensed atoms.
We solve Eq.~(\ref{eq:var}) in the case of harmonic motion
of the Gaussian sizes about their equilibrium values.
The frequencies of the lowest-lying quadrupole oscillation is \cite{perez-garcia96}
\begin{equation}
\begin{split}
\omega_Q& = \omega_r \sqrt{2} \bigg[ \left( 1 + \lambda^2 - P_{2,3} \right)  \\
       	& \quad - \sqrt{ \left( 1 - \lambda^2 + P_{2,3} \right)^2 
	+ 8 P_{3,2}^2 } \bigg]^{1/2},
\end{split}
\label{eq:quad}
\end{equation}
where $P_{i,j} = P / (4 u_{0r}^i u_{0z}^j)$ with
$u_{0z}$ and $u_{0r}$ the equilibrium axial and radial sizes, respectively.
%%%% added after review
The frequency of the $m = 0$ breathing mode
is the sum of the two terms in Eq.~\ref{eq:quad} rather than the difference,
and is a factor of about 60 higher in frequency for our experimental parameters.
%%%%%
For the case of a highly elongated trap ($\lambda \ll 1$),
in the Thomas-Fermi regime ($P \gg 1$), we find
the well known relation $\omega_Q = \omega_z\sqrt{5/2}$ \cite{stringari96,meanfield_correction},
and in the non-interacting limit ($P \rightarrow 0$),
we find $\omega_Q \rightarrow 2\;\omega_z$, as expected.

We may also use Eq.~(\ref{eq:var}) to determine the 
time evolution of the size of the BEC \cite{perez-garcia97}.
In particular, we are interested in the dynamics
associated with the modulation of $a$ using a Feshbach resonance,
which has been proposed previously
%\cite{adhikari1, PhysRevA.70.053604, abdullaev, PhysRevLett.90.230401, ramos:063412}
\cite{previousFRmanage,Adhikari2003211}.
In a magnetic field $B$, 
the scattering length near a Feshbach resonance may be described by
\begin{equation}
a (B) = a_{BG} \left( 1 - \frac{\Delta}{B - B_\infty} \right),
  \label{eq:feshbach}
\end{equation}
where $a_{BG}$ is the background scattering length,
$\Delta$ is the resonance width,
and $B_\infty$ is the resonance location.
We consider a time-dependent magnetic field of the form
\begin{equation}
B(t) = B_\mathrm{av} + \delta B \cos (\Omega t), \nonumber
\end{equation}
where $\Omega$ is the modulation frequency.
Using this form for $B$, 
the result of expanding Eq.~(\ref{eq:feshbach}) to first order 
in the small quantity $\delta B \ll |B_\mathrm{av} - B_\infty|$ is
\begin{equation}
a (t) \simeq a_\mathrm{av} + \delta a \cos (\Omega t),
\end{equation}
where
\begin{eqnarray}
 a_\mathrm{av} = a (B_\mathrm{av})
 & \mathrm{and} & 
 \delta a =  \frac{a_{BG} \,\Delta \, \delta B }{\left(B_\mathrm{av}-B_\infty\right)^2} \, . \nonumber
 \end{eqnarray}
This expression for $a$ is substituted into Eq.~(\ref{eq:var}),
and a fourth-order Runge-Kutta method is used to solve 
the system of equations numerically.  
The results may be directly compared with those from experiment.

%%%%%%%%%%%%%%%%%%%%%%%%%%%%%%%%%%%%%%%%%%%%%%%%%%%%%%%%%%%%%%%%%%%%%%%%%%%%%
%%%%%%%%%%%%%%%%%%%%%%%%%%%%%%%%%%%%%%%%%%%%%%%%%%%%%%%%%%%%%%%%%%%%%%%%%%%%%
%\section{Results and Discussion}
%%%%%%%%%%%%%%%%%%%%%%%%%%%%%%%%%%%%%%%%%%%%%%%%%%%%%%%%%%%%%%%%%%%%%%%%%%%%%

Our methods for creating an ultracold gas of $^7$Li have
been described previously \cite{Pollack,PollackScience}.
We confine atoms in the $|1,1\rangle$ hyperfine
state of $^7$Li in an optical trap and
use a set of Helmholtz coils to provide an axially oriented bias field.
We determine the radial trapping frequency by parametric heating
to be $\omega_r = (2\pi)\,235(10)\,\mathrm{Hz}$, and the
axial trapping frequency by dipole oscillation to be
$\omega_z = (2\pi)\,4.85(1)\,\mathrm{Hz}$.
After evaporation from the optical trap at 717\,G, where $a \sim 200\,a_0$,  
we have $N \sim 3\times10^5$ atoms with a condensate fraction $> 90\%$.
We then ramp the bias field in 4\,s from the initial value 
to $B_\mathrm{av} = 565$\,G (where $a_\mathrm{av} \sim 3\,a_0$).
For these experimental values, the dimensionality
parameter is $\lambda P \approx 0.3$, 
close to the transition into the quasi-1D regime 
(determined by $\lambda P~\ll~1$)~\cite{PhysRevA.66.043610}.
At this point we oscillate the magnetic field with a modulation 
depth of $\delta B = 14\,$G, corresponding to $\delta a \sim 2\,a_0$.
We use \emph{in situ} polarization phase-contrast
imaging to obtain the density distribution 
of the condensate~\cite{PhysRevLett.78.985}
to which we fit a Gaussian characterized by $1/e$ axial and radial half-widths.

In Fig.~\ref{fig:image}, we show pictures taken with $\Omega = (2\,\pi)\,10\,\mathrm{Hz}$.
A quadrupole oscillation of the cloud is readily observable.
%%% text added after review
The large oscillation amplitudes considered here
extend over approximately 1\,mm of the optical trap.
A harmonic approximation of the trapping potential about the trap center is
less than 10\% in error over this range.
The size of the cloud in Fig.~\ref{fig:image} as a function of time is modeled well by
the variational calculation, consistent with negligible anharmonic contributions.
Furthermore, we observe no damping of the quadrupole mode over many
oscillation periods, consistent with a negligible thermal fraction.
%%% 

%%%%%%%%%%%%%%%%%%%%%%%%%%%%%%%%%%%%%%%%%%%%%%%%%%%%%%%%%%%%%%%%%%%%%%%%
\begin{figure}
\includegraphics[height=1\columnwidth,angle=-90]{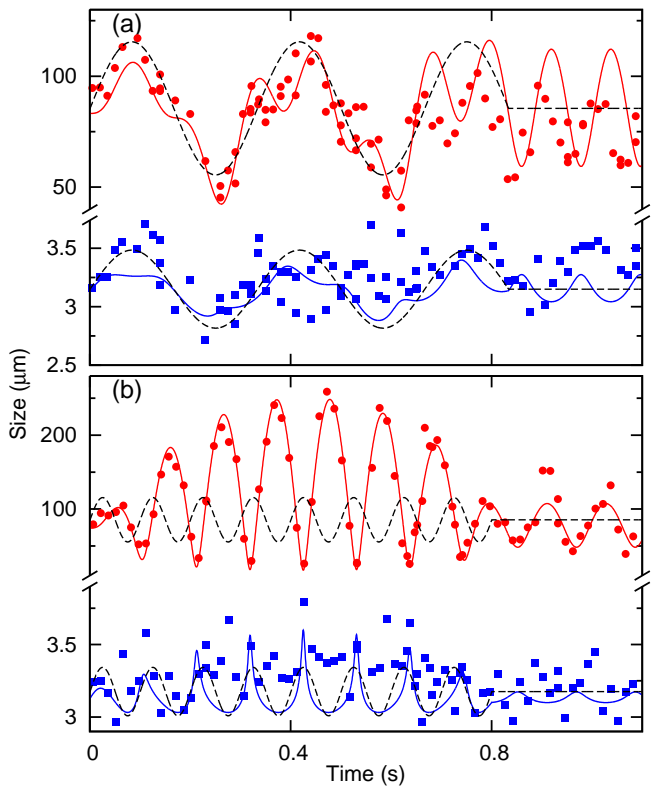}
\caption{ (color online)
    Axial (red circles) and radial (blue squares) $1/e$ sizes of a condensate 
    during and after excitation 
    with $a_\mathrm{av} \sim 3\,a_0$ and $\delta a \sim 2 \,a_0$, where
    (a) $\Omega = (2\pi)\,3\,$Hz or (b) $\Omega = (2\pi)\,10\,$Hz.
    The natural oscillation frequency is $\omega_Q = (2\pi)\,8.2(1)\,\mathrm{Hz}$.
    Note that the 1.8\,Hz beat note is the cause of the diminished amplitude in (b)
    after the modulation drive is turned off.
    The solid lines are results from the variational calculation 
    using the same parameters as in the experiment and contain no adjustable parameters.
    The dashed lines are representative of the driving excitation.
    The resolution of the optical imaging system is $\sim$3\,$\mu$m, which
    limits accurate determination of the radial sizes.
    The frequency of the breathing mode is on order 470\,Hz and therefore
    no effects of this mode are present.
\label{fig:driveFree} }
\end{figure}
%%%%%%%%%%%%%%%%%%%%%%%%%%%%%%%%%%%%%%%%%%%%%%%%%%%%%%%%%%%%%%%%%%%%%%%%

Results from the variational calculation show that during the excitation
if $\Omega < \omega_Q$, then the axial and radial sizes of the cloud 
follow the change in $a$: growing as $a$ increases, and shrinking
as $a$ decreases.
%%%%% added after review
This in-phase behavior of both the axial and radial sizes of the 
cloud is expected for adiabatic changes in $a$,
and therefore 
should not be confused with the high-lying $m=0$ breathing mode
for low frequencies.
%%%%%
However when $\Omega > \omega_Q$, the radial size follows
nearly in-phase, while the axial size lags behind the radial size 
by half a period---an out-of-phase oscillation.  
In both cases when the excitation
is stopped, the cloud undergoes free quadrupole oscillations with the
axial and radial sizes $\pi$ out-of-phase.
Characteristic data for these two regimes are shown in Fig.~\ref{fig:driveFree}
and reasonably agree with the variational results.
We fit the time evolution of the size of the cloud after excitation 
and determine the free quadrupolar oscillation frequency to be $\omega_Q = (2\pi)\,8.2(1)\,$Hz,
in good agreement with the predicted value of 8.17\,Hz from Eq.~\ref{eq:quad},
where we have used $\lambda = 0.021$ and $P \approx 15$.
%%%% added after review
Similar agreement between measured and predicted
quadrupole frequencies in the dimensional crossover regime have
been previously observed \cite{PhysRevA.72.053631}.
%%%%
The amplitude of this free oscillation is dependent on the duration
of excitation as well as the phase of the driving force at the time when the 
excitation is stopped.  Therefore, care must be taken when comparing data with theoretical
predictions of the amplitude during the free oscillation.

\begin{figure}
\includegraphics[height=1\columnwidth,angle=-90]{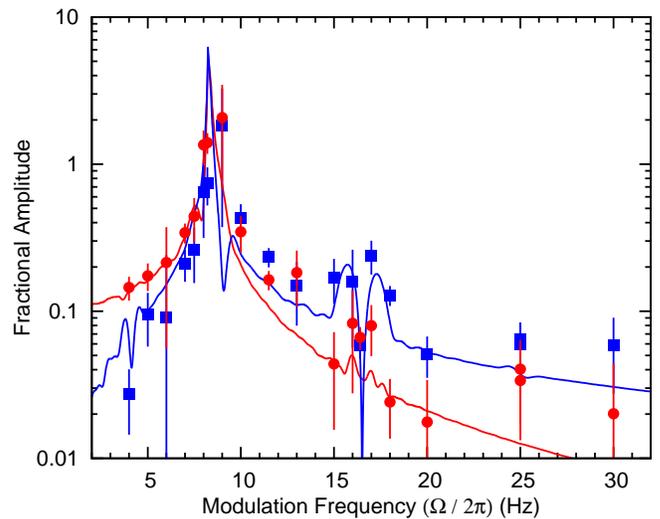}
\caption{ (color online)
    Fractional amplitude of the drive (red circles) 
    and quadrupole (blue squares) 
    modes during excitation at frequency $\Omega$
    with a modulation depth $\delta a \sim 1\,a_0$, where $a_\mathrm{av} \sim 3\,a_0$.
    The solid lines are results from the variational calculation with no adjustable parameters.
    %The fractional amplitude is defined as the axial amplitude of the oscillation 
    %scaled by the equilibrium $1/e$ axial size of the condensate. 
    The oscillation is notably asymmetric for fractional amplitudes of order 1 and larger,
	as shown in Fig.~\ref{fig:driveFree}(b).
\label{fig:spectrum} }
\end{figure}

A less parameter-dependent measure is to observe the amplitude of
the oscillation \emph{during} excitation.  
As can be discerned from Fig.~\ref{fig:driveFree},
during excitation the size of the condensate oscillates at
both the drive frequency and the natural quadrupole frequency.
Beating between these frequencies modulates the instantaneous deviation from the unperturbed size.
An ideal method for determining the energy in the 
driven mode and the excited quadrupole mode 
separately is to use Fourier analysis~\cite{PhysRevA.56.4855}.  
This method can be experimentally difficult, however, 
given the required number of 
coherent oscillations needed to accurately resolve the sinusoidal
peak in the Fourier spectrum.  
Instead, we assume that the system 
can be described by the linear combination of
sinusoids at the known frequencies $\Omega$ and $\omega_Q$.
%and fit the observed condensate sizes.
After driving an excitation for 0.5\,s, we 
%extract the amplitude of both spectral components 
fit the observed condensate axial size to the following expression 
during an additional 0.5\,s of excitation:
\begin{equation}\label{eqn:fit}
u(t) = u_0 + 
u_\Omega \sin \left( \Omega\, t + \Phi \right) +
u_Q \sin \left( \omega_Q t + \phi \right),
\end{equation}
where $u_0$ is the equilibrium $1/e$ size, $u_\Omega$ and $u_Q$ are
the amplitudes of the drive and quadrupole modes, respectively,
and $\Phi$ and $\phi$ are the respective phases.
The fractional amplitudes $u_\Omega/u_0$ and $u_Q/u_0$ are
shown as functions of $\Omega$ in Fig.~\ref{fig:spectrum}.
%are the fitted axial amplitudes of the oscillations 
%at $\Omega$ and $\omega_Q$ scaled by the unperturbed 
%axial size of the condensate.
As expected, there is a resonant enhancement
in both the quadrupole and drive modes when $\Omega = \omega_Q$.  
In addition, we see a parametric enhancement when $\Omega \approx 2\,\omega_Q$. 
The dip at exactly $2\,\omega_Q$ is due to
destructive interference between the drive mode and the resonantly excited mode.
This interference structure is observed in both the data and the simulation.

%%%%%%%%%%%%%%%%%%%%%%%%%%%%%%%%%%%%%%%%%%%%%%%%%%%%%%%%%%%%%%%%%%%%%%%%
%%%%%%%%%%%%%%%%%%%%%%%%%%%%%%%%%%%%%%%%%%%%%%%%%%%%%%%%%%%%%%%%%%%%%%%%

%%%%%%%%%%%%%%%%%%%%%%%%%%%%%%%%%%%%%%%%%%%%%%%%%%%%%%%%%%%%%%%%%%%%%%%%
\begin{figure}
\includegraphics[height=1\columnwidth,angle=-90]{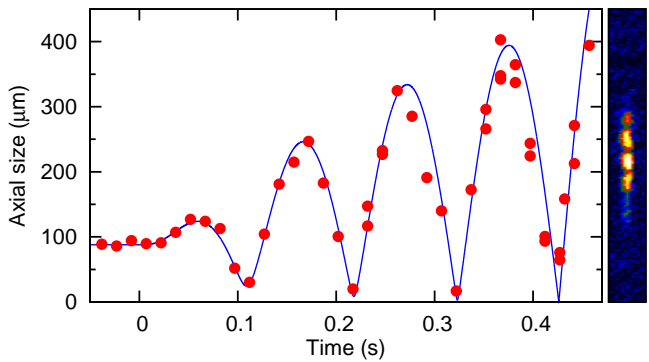}
\caption{ (color online)
    Quadrupole oscillation driven by a large amplitude excitation,
    $\delta a \sim 3 a_0$ where $a_\mathrm{av} \sim 3\,a_0$, 
    near resonance $\Omega = (2\pi)\,9\,$Hz.
    The non-sinusoidal behavior leads to fragmentation
    of the condensate during the compression stages
    when the axial size becomes smaller than the 
    axial harmonic oscillator size of $\sim$17\,$\mu$m.
    The image at right was taken during the compression stage at $t = 0.52\,$s 
    and shows fragmentation of the condensate,
    the field of view is $25\,\mu\mbox{m} \times 380\,\mu\mbox{m}$.
\label{fig:largeAmplitude} }
\end{figure}
%%%%%%%%%%%%%%%%%%%%%%%%%%%%%%%%%%%%%%%%%%%%%%%%%%%%%%%%%%%%%%%%%%%%%%%%
%%%%%%%%%%%%%%%%%%%%%%%%%%%%%%%%%%%%%%%%%%%%%%%%%%%%%%%%%%%%%%%%%%%%%%%%
\begin{figure}
\includegraphics[height=1\columnwidth,angle=-90]{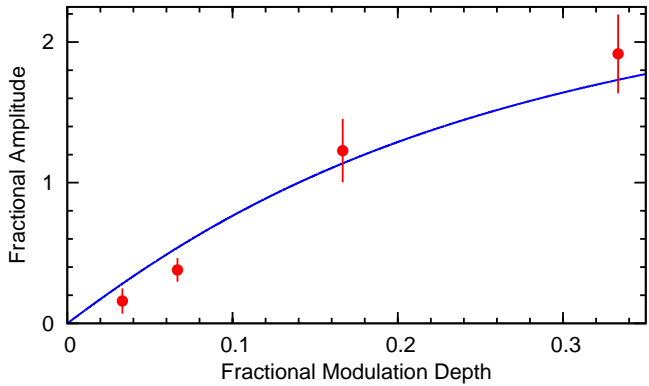}
\caption{Fractional amplitude of oscillation, when $\Omega \approx \omega_Q$,
as a function of the fractional modulation depth $\delta a / a_\mathrm{av}$ 
after 1\,s of excitation.
The solid curve is the result from the variational calculation.
\label{fig:depthPeak} }
\end{figure}
%%%%%%%%%%%%%%%%%%%%%%%%%%%%%%%%%%%%%%%%%%%%%%%%%%%%%%%%%%%%%%%%%%%%%%%%

Larger drive amplitudes push the oscillations into the nonlinear regime
where the amplitude of oscillation is no longer linearly dependent on
the modulation depth.
The first noticeable effect is the non-sinusoidal 
behavior of the oscillation seen in Fig.~\ref{fig:driveFree}(b).
In Fig.~\ref{fig:largeAmplitude} we show the result of driving the system
near resonance with a fractional modulation depth $\delta a/a_\mathrm{av} = 1$.
As the amplitude of the driven oscillation grows, it 
eventually becomes comparable with the original condensate size.
At this point, the size of the condensate cannot become smaller,
and therefore, the oscillation becomes increasingly asymmetric.  
In this manner, the rate of energy transfer into the oscillation will decrease.  
At even larger amplitudes, we observe that the condensate appears to fragment \cite{Adhikari2003211}.  
Similar looking results have been observed when modulating the radial confinement of a
cigar-shaped BEC, where Faraday waves may be excited~\cite{PhysRevLett.98.095301}.
By fitting the excitation spectrum (Fig.~\ref{fig:spectrum}) to a skew Lorentzian
(where the peak excitation frequency, width, amplitude, and skewness are fit parameters),
we determine the maximum fractional amplitude of oscillation of the quadrupole mode
after 1\,s of excitation.  
Our experimental results for the amplitude as a function
of modulation depth are presented in Fig.~\ref{fig:depthPeak}
along with results from the variational calculation, which
show good agreement with no adjustable parameters.
%%%% text added after review
We note that for the smallest of modulation depths investigated,
we only observed oscillation of the condensate when the drive
frequency was near resonance.  In addition, we found a roughly
linear scaling of the instantaneous amplitude of the 
quadrupole oscillation with the duration of the excitation in this regime. 
Furthermore, Fig.~\ref{fig:depthPeak} conveys to us that the data in
Fig.~\ref{fig:driveFree}, which was driven
at a fractional modulation depth of $\sim$0.7, 
was not in the linear regime,
and the data shown in Fig.~\ref{fig:spectrum},
driven at a fractional modulation depth of $\sim$0.3,
had a response that departed from linearity by about 30\%.
%%%%%

Large amplitude oscillations are accompanied 
by a frequency shift of the quadrupole mode~\cite{Pitaevskii1997406}.
For our geometry, this shift can
be approximated by $\delta \omega / \omega_Q \approx 0.1 A_z^2$ for small $A_z$,
where $A_z \equiv u_Q/u_0$ is the fractional amplitude of the axial size~\cite{PhysRevA.56.4855}.
The data shown in Fig.~\ref{fig:spectrum} has $A_z \sim 2$ on resonance,
and therefore in this case the above approximation for the frequency shift is not valid.
Our simulated results show a frequency shift $\sim$8\% for $A_z = 2$ with
a shallow amplitude dependence at larger $A_z$.  
Whereas a shift of 10\% was observed in the oscillation of Fig.~\ref{fig:largeAmplitude}, 
%Although a frequency shift of 10\% is in principle measurable for our data 
we were not able to resolve frequency shifts by fitting Lorentzians 
to the excitation spectra
for our data at low $A_z$ shown in Fig.~\ref{fig:depthPeak}.
There is an additional frequency shift due to 
non-zero temperature \cite{jin97,stamper-kurn98,PhysRevLett.88.250402,PhysRevA.72.053631}
which is estimated to be negligible given our low
temperatures and interaction strength \cite{PhysRevA.61.063615}.
Even though we can neglect temperature effects in the measurements presented here,
our method of excitation of the quadrupole mode
may be used to study these effects in further detail in regimes 
of stronger interactions.
In addition, going to large values of $a$ will facilitate investigations of
beyond mean-field effects on the collective modes of a Bose gas~\cite{pitaevskii98},
complementary to those observed in a Fermi gas at the BCS-BEC crossover~\cite{PhysRevLett.92.150402,altmeyer:040401,riedl:053609}.

%%%%%%%%%%%%%%%%%%%%%%%%%%%%%%%%%%%%%%%%%%%%%%%%%%%%%%%%%%%%%%%%%%%%%%%%%%%%%
%\section{Conclusion}
In this work, we have experimentally demonstrated the excitation of
the collective low-lying quadrupole mode of a dilute Bose gas
by modulating the atomic scattering length.  Our observations
are supported by variational calculations of the
time dependent Gross-Pitaevskii equation assuming
a Gaussian trial wavefunction.
Using this formalism we find good agreement with our experimental results.
Temporal modulation of the scattering length,
as afforded by Feshbach resonances, provides an additional tool
for exciting collective modes of an ultracold
atomic gas.  
This method is quite attractive in circumstances
where excitation of the condensate by other means, such
as trap deformation, is unavailable.  
In addition, this method can be used for
condensates in the presence of thermal atoms
where principal excitation of the condensate alone is desired,
as well as in multi-component gases 
where excitation of only one species can be accomplished.

%%%%%%%%%%%%%%%%%%%%%%%%%%%%%%%%%%%%%%%%%%%%%%%%%%%%%%%%%%%%%%%%%%%%%%%%%%
%\begin{acknowledgments}
We thank V. I. Yukalov for stimulating discussions related to this work.
The work at Rice was funded by the NSF, ONR, 
the Keck Foundation, and the Welch Foundation (C-1133).
The work at S\~{a}o Carlos was funded by
FAPESP~-~CEPID, CNPq/FAPESP~-~INCT and CAPES.
%\end{acknowledgments}

\bibliography{../bibliography}

\begin{thebibliography}{10}
\bibitem{meanfield_correction}
The Bogoliubov theory can be used to determine a quantum correction to
  Eq.~\ref{eq:quad} which is proportional to $(N a^3 / u_r^2 u_z)^{1/2}$
  \cite{pitaevskii98}. For our experimental parameters this term is negligible.
\end{thebibliography}

\end{document}